\documentclass[12pt]{iopart}
\usepackage{epsfig,amssymb}
\def\beq{\begin{equation}}
\def\eeq{\end{equation}}
\def\bea{\begin{eqnarray}}
\def\eea{\end{eqnarray}}
\def\d{{\mathrm{d}}}

\newfont{\cursive}{pzcmi at 9pt}
\def\~t{\tilde{t}}
\def\Painleve{Painlev\'{e}}

\def\e2phi{\e^{2\Phi}}
\begin{document}
\title{Dynamical surface gravity.}
\author{Alex B Nielsen $^1$ and Jong Hyuk Yoon $^2$}
\address{$^1$Department of Physics and Astronomy, University of Canterbury,
Private Bag 4800, Christchurch, New Zealand \footnote{Current address: Center for Theoretical Physics, Seoul National University, Seoul 151-747, Korea}}

\address{$^2$Department of Physics, Konkuk University, Seoul 143-701, Korea}
\eads{\mailto{eujin@phya.snu.ac.kr}, \;\mailto{yoonjh@konkuk.ac.kr}}
\begin{abstract} We discuss how the surface gravity can be classically defined for dynamical black holes. In particular we focus on defining the surface gravity for locally defined horizons and compare a number definitions proposed in the literature. We illustrate the differences between the various proposals in the case of an arbitrary dynamical, spherically symmetric black hole spacetime. We also discuss how the trapping horizon formalism of Hayward can be related to other constructions.
\end{abstract}
%
%
\section{Introduction}

In black hole thermodynamics the surface gravity of a black hole plays a role
analogous to temperature.  The relationship between black hole evolution and thermodynamics is one of the most widely studied topics in theoretical physics. Indeed the discovery of the semi-classical Hawking radiation effect put black hole thermodynamics on a firm theoretical footing and directly established the relationship between the surface gravity and the temperature. However, the derivation of the Hawking radiation effect depends on quasi-stationary, quasi-equilibrium evolution. In a non-equilibrium, dynamical
situation the temperature may not be well defined. 

In a fully dynamical situation, the surface gravity will probably not be directly analogous to a temperature of any thermal spectrum. However, there are two reasons for investigating the surface gravity in dynamical situations. The first is the desire to derive a purely classical evolution law for the black hole, in the sense of the first law, without making any appeal to true thermal behaviour. In this context the surface gravity will play the role as the `constant' of proportionality between the change in the mass of the black hole and the change in the area. In this way the definition of the surface gravity is closely connected to our choice of quasi-local mass for the black hole.

The second reason is that the surface gravity is likely to play a key role in the emission of Hawking radiation, even in non-equilibrium processes. There have been a number of derivations of the Hawking effect that are easily applicable to dynamical situations (see for example \cite{Hajicek:1986hn,Parikh:1999mf,Visser:2001kq}). These derivations make it clear that the Hawking effect is a local geometrical effect and that something resembling the surface gravity plays a key role. Heuristically, one is tempted to think that the larger the surface gravity, the more Hawking radiation there will be and the quicker the black hole will evaporate.

The most widely known definition of the surface gravity is in terms of a Killing horizon. This is for example, how the surface gravity is calculated for a Schwarzschild black hole. This works well in stationary situations but breaks down in fully dynamical situations, where no such Killing horizon exists. A key question is whether the surface gravity can be defined for black holes that are evolving, either by accreting matter or by emitting Hawking radiation.

In stationary spacetimes the event horizon of a black hole is typically a Killing horizon for a suitably chosen Killing vector. However, for event horizons in more general, non-stationary spacetimes, the surface gravity can be defined in terms of the null generators of the horizon. In these more general cases there is no Killing vector field to use to fix the normalization of the surface gravity. Since the generator of the horizon is only really defined on the event horizon, there is no natural way of fixing this normalization by imposing a condition off the horizon.

However, recently much interest has focused on local definitions of horizons \cite{Hayward:1993wb,Ashtekar:2004cn,Nielsen:2007lh}. These local horizons are typically defined in terms of spacelike two-surfaces for which the expansion of the outgoing null normal vanishes. Indeed, there are good reasons for believing that it is such a local horizon, and not the event horizon, that is responsible for the Hawking process \cite{Hajicek:1986hn,Visser:2001kq}. Since we expect the surface gravity to play a role in governing the `amount' of Hawking radiation, it makes sense to investigate surface gravity definitions for local horizons in dynamical situations.

While there have been many proposals for what the dynamical surface gravity should be, there is not yet any consensus as to what the correct definition is. In this paper, we investigate a number of different definitions for the dynamical surface gravity that have appeared in the literature. In order to compare them, we will compute their forms for marginally trapped surfaces in a generic, dynamic, spherically symmetric spacetime.  Furthermore, we will be able to evaluate how these definitions compare to the familiar Killing horizon definition in
the static limit.

The results presented here will be of a purely classical nature based purely on the geometry of the spacetime. We will not perform any quasi-classical calculations involving quantum fields and we will not prove any relation between the classical surface gravity discussed here and any Hawking radiation that may be emitted, although such a relation is of course one of the main motivations for the work. We rather take the viewpoint that the surface gravity can be defined in a classical way and thus its nature is tied up with the local geometrical properties of the space-time at the horizon.

The paper is organized as follows. In section \ref{section:surface gravities} we give an overview of the various proposals that have appeared in the literature for defining surface gravity including for Killing horizons and event horizons. In section \ref{section:spherically symmetric} we go on to compare these various definitions in the case of a fully general, dynamical, spherically symmetric spacetime. We then conclude with a discussion.

When working in advanced Eddington-Finkelstein coordinates we will use a dot to denote differentiation with respect to the time coordinate $v$, $\dot{m} = \partial_{v}m$, and a dash to denote differentiation with respect to the radial coordinate $r$, $m'=\partial_{r}m$. Greek superscripts and subscripts denote components in a given coordinate basis, while Latin superscripts and subscripts denote tensor indices in the abstract index notation \cite{Wald:book}.

\section{Surface gravity definitions}

\label{section:surface gravities}

\subsection{Surface gravity for Killing horizons and event horizons}

The traditional definition of surface gravity \cite{Bardeen:1973gs}, is based on the idea of a Killing horizon, a hypersurface where a Killing vector of the metric becomes null. In general relativity, stationary event horizons are typically Killing horizons for a suitably chosen Killing vector $k^{a}$ (for the technical assumptions for this proposition see \cite{Chrusciel:1996bm}). The surface gravity $\kappa$ of the Killing horizon can be defined by
\beq \label{killgeodes} k^{a}\nabla_{a}k^{b} = \kappa k^{b}. \eeq
Thus, the surface gravity for a Killing horizon is defined by the fact  that the Killing vector is a non-affinely
parameterized geodesic on the Killing horizon\footnote{In general, off the horizon, the Killing vector is not geodesic.}. The proof of this is as follows: Since $k^{a}k_{a}$ is constant and zero on the Killing horizon by definition, then $\nabla_{b}(k^{a}k_{a})$ must be normal to the horizon, in the sense of being orthogonal to any vector tangent to a curve lying in the horizon. Since $k^{a}$ is also normal to the horizon, and this normal is unique, we have $k^{a}\nabla_{b}k_{a}\propto k_{b}$. Using the Killing relation, we then obtain the above.

We note here, for later use, that due to Killing's equation the above can also be written as
\beq \label{killversion2} \frac{1}{2}g^{ab}k^{c}\left(\nabla_{c}k_{a} - \nabla_{a}k_{c}\right) = \kappa k^{b}. \eeq
%
Since $k^{a}$ is normal to the null hypersurface, it is hypersurface orthogonal and by the Fr\"{o}benius theorem, we have
\beq \label{hypsurcond} k_{[a}\nabla_{b}k_{c]} = 0. \eeq
Contracting this condition (\ref{hypsurcond}) with $\nabla^{a}k^{b}$ and using the Killing property we get another formula often seen in the literature as a definition of surface gravity \cite{Wald:book},
\beq \label{kappasqrd} \kappa^{2} =
-\frac{1}{2}(\nabla^{a}k^{b})(\nabla_{a}k_{b}). \eeq
This version also provides us with a nice physical interpretation of the surface gravity. In static spacetimes the surface gravity can be interpreted as the limiting force required at infinity to hold a mass stationary above the event horizon (one could imagine the mass connected to infinity by a very long massless string). Note that this physical interpretation does not work in non-static spacetimes \cite{Wald:book}.

By virtue of Killing's equation $\nabla_{a}k_{b}+\nabla_{b}k_{a}=0$, the Killing vector is only determined up to a constant factor. This freedom corresponds to a gauge freedom to rescale the curve parameter along the integral curves of the Killing vector by a constant factor. The advantage of using a Killing horizon of a global Killing vector field is that for static, asymptotically flat spacetimes, this factor can be fixed by requiring $k_{a}k^{a}=-1$ at infinity and thus, at infinity, the Killing vector coincides with the four-velocity of a static observer parameterized by the observers proper time $\tau$.\footnote{A slightly different prescription is required for non-static spacetimes such as the Kerr solution, since the relevant Killing vector is no longer composed purely of a time-translational Killing vector. In this paper we are only considering spherically symmetric spacetimes.} This fixes the freedom in $k^{a}$ globally and uniquely determines the value of the surface gravity of a Killing horizon.

This prescription can also be extended to apply to static observers anywhere in the spacetime and leads to the idea that the temperature of the event horizon tends to infinity as seen by static observers located closer and closer to the horizon. However, this procedure relies on a Killing vector field to relate the normalization to a point away from the horizon. 

A similar definition to (\ref{killgeodes}) can be made for an event horizon that is not a Killing horizon, with the role of the Killing vector replaced by the null generator of the event horizon. However, in the general case, there is no natural way to fix the parametrization of this generator. Even in spherically symmetric spacetimes, where the generators of the future event horizon will form a subset of the congruence of radially outgoing null geodesics it is not easy to fix the parameterization of the generators and thus the value of the surface gravity.
\subsection{Surface gravity for marginally trapped surfaces}

In dynamical situations, it is a well known fact that local definitions of horizons such as apparent horizons, trapping horizons or dynamical horizons do not necessarily coincide with the location of the event horizon\footnote{This issue is actually quite subtle. In spacetimes satisfying the Null Energy Condition, it has been conjectured by Eardley \cite{Eardley:1997hk} that the outer boundary of the region containing marginally trapped surfaces is in fact the event horizon. However, in spherically symmetric spacetimes, with spherically symmetric slicings, the apparent horizon is in most cases also a dynamical horizon and a trapping horizon. In most dynamical cases it does not coincide with the location of the event horizon \cite{Nielsen:2005af}.}. In such cases one is left with the question `for which surface should one define the black hole area or the black hole surface gravity?' The canonical choice is to use the event horizon. However, as noted above, there is evidence that it is the apparent horizon, and not the event horizon, that plays the key role in Hawking radiation. This idea has become a key point in hopes to demonstrate Hawking radiation in the laboratory using analogue gravity models \cite{Barcelo:2005fc}. Furthermore, it has been shown that the laws of black hole thermodynamics can easily be applied to the case of trapping \cite{Hayward:1993wb} and dynamical \cite{Ashtekar:2004cn} horizons.

Most of the definitions we will consider in this work are motivated by investigations of local horizons. Therefore we will evaluate the surface gravity for marginally outer trapped surfaces, although several of the definitions could equally well be applied to event horizons.

A marginally trapped surface is a compact spacelike two-surface for which the expansion of one of the future-directed null normals vanishes. In the following we will consider spacelike two-surfaces with an ingoing null normal $n^{a}$ and an outgoing null normal $l^{a}$. Furthermore, we will assume that the cross-normalization of these two null vectors is such that $l^{a}n_{a}=-1$. A marginally outer trapped surface is therefore defined by the requirement that $\theta_{l} = 0$ where the expansion is given by
\beq \label{expanl} \theta_{l} = g^{ab}\nabla_{a}l_{b} +
n^{a}l^{b}\nabla_{a}l_{b} + l^{a}n^{b}\nabla_{a}l_{b}. \eeq
A common approach to defining the surface gravity of a non-Killing horizon is to use the fact that $l^{a}$ is typically a non-affinely parameterized geodesic on the horizon, although \emph{it is not always a horizon generator}. This mirrors the requirement for a Killing horizon that the Killing vector should be a non-affinely parameterized null geodesic on the Killing horizon.
In this way the surface gravity can be defined via the equation
\beq l^{a}\nabla_{a}l^{b} = \kappa l^{b}, \eeq
or
\beq \kappa = -n^{b}l^{a}\nabla_{a}l_{b}. \eeq
While the direction of the null geodesic $l^{a}$ will be fixed by the location of the horizon and its foliations into spacelike two-spheres, the choice of parametrization of $l^{a}$ will become crucial to the overall value of the surface gravity.

Writing the null vector $l^{a}$ as the tangent vector of a curve $x^{\mu}(\lambda)$ with parameter $\lambda$, under a change of parametrization of the curve $\lambda \rightarrow \lambda'$, the components of the tangent vector change by\footnote{Note this change of parametrization can depend on the spacetime point $x$.}
\beq l^{\mu} \equiv \frac{\d x^{\mu}}{\d\lambda} \rightarrow l^{\mu'} \equiv
\frac{\d x^{\mu}}{\d\lambda'} =
\frac{\d\lambda}{\d\lambda'}(x)l^{\mu} = \Omega(x) l^{\mu}, \eeq
and the surface gravity changes by
\beq \label{changesurfgrad} \kappa \rightarrow \kappa' =
\frac{\d\lambda}{\d\lambda'}(x)\kappa +
l^{a}\nabla_{a}\left(\frac{\d\lambda}{\d\lambda'}\right)(x) = \Omega(x)\kappa + l^{a}\nabla_{a}\Omega(x). \eeq
If the surface gravity is to measure the extent to which the geodesic $l^{a}$ fails to be affinely parameterized, the choice of parametrization becomes essential. For a Killing horizon the normalization is fixed by the link between the parametrization of the Killing vector and the proper time at infinity.

For an affinely parameterized geodesic $\kappa = 0$. In general the tangent vector $l^{a}$ can always be reparameterized to eliminate $\kappa$. However, a number of proposals have been made to fix the parameterization of this null vector and in order to give a non-zero surface gravity this parameterization is required to be non-affine. The following are some of the proposals that have been given for fixing this normalization.\bigskip

Fodor et al. \cite{Fodor:1996rf} proposed a non-local choice of normalization for spherically symmetric spacetimes in terms of an affinely-parameterized ingoing null geodesic $n^{a}$ whose asymptotic form was such that $t^{a}n_{a}=-1$ where $t^{a}$ is the asymptotic time-translational Killing vector. In order to do this one needs to require that the spacetime admits an asymptotically flat spatial infinity. This choice of parameterization for ingoing $n^{a}$ is then transferred to the choice of parameterization of outgoing $l^{a}$ by requiring the condition $l^{a}n_{a}=-1$ everywhere in the spacetime. The proposal of \cite{Fodor:1996rf} is therefore
\beq \kappa_{F} = -n^{b}l^{a}\nabla_{a}l_{b}, \eeq
where $n^{a}$ must be affinely parameterized everywhere and at asymptotically flat spatial infinity by the proper time of static observers. Fodor et al. discuss how this normalization can be observed locally by measuring the frequency of fiducial photons sent in from infinity.\bigskip

Hayward \cite{Hayward:1997jp} gave a definition of surface gravity for dynamic, spherically symmetric spacetimes in terms of the Kodama vector $K^{a}$ \cite{Kodama:1979vn}. The Kodama vector has the property that the combination $K_{a}T^{ab}$ is divergence free in spherical symmetry and that $K^{a}$ reduces to $K^{a}K_{a}=-1$ at spatial infinity. The overall sign can be fixed by requiring it to be future directed. In static electrovac solutions it coincides with the time-translational Killing vector of the Reissner-Nordstr\"{o}m geometry \cite{Hayward:1997jp}. The value of the `dynamic surface gravity' for a trapping horizon was defined by
\beq \frac{1}{2}g^{ab}K^{c}\left(\nabla_{c}K_{a}- \nabla_{a}K_{c}\right) = \kappa_{\mathrm{Ko}} K^{b}. \eeq
This matches the form of (\ref{killversion2}) but note that the Kodama vector does not in general satisfy Killing's equation and it is not necessarily geodesic on the horizon. The surface gravity defined in this way is unique since $K^{a}$ is unique and agrees with the surface gravity in the Reissner-Nordstr\"{o}m case. However, this definition does not agree with that given in \cite{Hayward:1993wb} or \cite{Fodor:1996rf} and is only applicable in spherically symmetric spacetimes.

For an isolated horizon, Ashtekar, Beetle and Fairhurst \cite{Ashtekar:1999yj} fixed the normalization by setting the expansion of the ingoing null vector $n^{a}$ to be the same as the Reissner-Nordstr\"{o}m value and cross-normalizing with $l^{a}$ via $n^{a}l_{a}=-1$. This idea was elaborated by Ashtekar, Fairhurst and Krishnan \cite{Ashtekar:2000hw} by fixing $\kappa$ as a unique function of the horizon parameters $a_{\triangle}$ and $Q_{\triangle}$, in terms of the known value given in the static case. However it is not possible to extend this method to situations such as Einstein-Yang-Mills where the surface gravity in the static sector is not a unique function of the horizon parameters.

In the case of a slowly evolving horizon, defined in terms of a perturbative expansion around an isolated horizon, Booth and Fairhurst \cite{Booth:2003ji} gave a new definition of surface gravity that was an extension of the definition for isolated horizons. It is given by
\beq \kappa_{\mathrm{B}} = -Bn^{a}l^{b}\nabla_{b}l_{a}-Cl^{a}n^{b}\nabla_{b}n_{a}, \eeq
where the normal to horizon is given by $\tau^{a} = Bl^{a}+Cn^{a}$. $B$ and $C$ are just scalar fields on the dynamical horizon that encode how the choice of a horizon normal (unique up to a factor) can be expressed in terms of the chosen $l^{a}$ and $n^{a}$. Clearly in the isolated horizon case, where $B=1$ and $C=0$, it reduces to the original isolated horizon definition. In addition, it is clear in this form that this definition incorporates information about the inaffinity of $l^{a}$, through the term $n^{a}l^{b}\nabla_{b}l_{a}$, and the inaffinity of $n^{a}$, through the term $l^{a}n^{b}\nabla_{b}n_{a}$.

It is also possible to define a surface gravity without appealing to the inaffinity of a null vector. This is usually done by identifying the surface gravity in a dynamical law for the black hole evolution. An analogue of surface gravity, called \emph{trapping gravity}, was presented by Hayward in \cite{Hayward:1993wb}. In the notation used here it is given by
\beq \kappa_{\mathrm{H}} = \frac{1}{2}\sqrt{-n^{a}\nabla_{a}\theta_{l}}. \eeq
On a marginally outer trapped surface, where $\theta_{l}=0$ this definition is independent of the parametrization of $l^{a}$ since under a reparameterization of $l^{a}$ by $\lambda\rightarrow\lambda'$, we have $l^{a}\rightarrow\frac{\d\lambda}{\d\lambda'}l^{a}$, $n^{a}\rightarrow\frac{\d\lambda'}{\d\lambda}n^{a}$ (since we require $n^{a}l_{a}=-1$) and $\theta_{l}\rightarrow\frac{\d\lambda}{\d\lambda'}\theta_{l}$. 

Another definition of surface gravity for a trapping horizon was proposed by Mukohyama and Hayward \cite{Mukohyama:1999sp}
\beq \kappa_{\mathrm{M}} = -\frac{1}{16\pi r}\int_{S}\d^{2}\theta\sqrt{h}\left(n^{a}\nabla_{a}\theta_{l}+l^{a}\nabla_{a}\theta_{n}+\frac{\theta_{n}}{\Lambda}l^{a}\nabla_{a}\Lambda \right). \eeq
%
%
Their original definition was also based on affinely parameterized null vectors. However, the original definition also included a factor of $l^{a}n_{a}$ which will not be $-1$ if both null vectors are affinely parameterized. Here we have introduced a factor $1/\Lambda$ that would make a general $l^{a}$ affinely parameterized. A full derivation of this result appears in \ref{doublenull}. Since this formula involves an integral over the marginally trapped surface it is manifestly quasi-local. Once again this definition is independent of the parametrization of $l^{a}$, provided the normalization $n^{a}l_{a}=-1$ is preserved.

For a dynamical horizon \cite{Ashtekar:2004cn}, the \emph{effective surface gravity} is identified from an area balance law \cite{Ashtekar:2004cn} and takes the value
\beq \frac{1}{2r}\frac{\d f(r))}{\d r} \eeq
where $r$ is the area defining radial coordinate and $f(r)$ is any function of $r$. Once again there is freedom in the normalization that can usually be fixed by appeal to the stationary Kerr solution.  

\section{Surface Gravity in dynamic, spherically symmetric spacetimes}

\label{section:spherically symmetric}

To see how these various definitions compare to one another we turn now to a specific example of a dynamical, spherically symmetric metric. A general, dynamic, spherically symmetric metric can be written in advanced Eddington-Finkelstein coordinates\footnote{These coordinates will be well-defined on the future horizon (but not the past horizon) and are therefore better suited to examining functions defined there than Schwarzschild coordinates. \Painleve-Gullstrand coordinates served a similar purpose in \cite{Nielsen:2005af}.} as
\beq \d s^{2} = - A^{2}(v,r)\triangle(v,r)\d v^{2} + 2A(v,r)\d v\d r + r^{2}\d\Omega^{2}, \eeq
The two free functions $A(v,r)$ and $\triangle(v,r)$ could in principle be determined by solving the full Einstein equations, although we will see here that this is not necessary for our purposes. The function $\triangle(v,r)$ can be written in terms of the Misner-Sharp mass function $m(v,r)$ as
\beq \triangle(v,r) = 1-\frac{2m(v,r)}{r}. \eeq
The function $A(v,r)$ will also play an important role in what follows and should not be confused with the area of the horizon.
%
%
Note that in, general, the function $A(v,r)$ cannot be gauged away and a general spherically symmetric spacetime must contain two free functions \cite{Nielsen:2005af}. In certain situations, such as the Schwarzschild and Reissner-Nordstr\"{o}m solutions, one can consistently choose $A=1$. However, in other situations, such as the Einstein-Skyrme system studied in \cite{Nielsen:2006gb}, this function cannot be set equal to one everywhere and its value on the horizon will affect the value of the surface gravity computed in terms of a Killing horizon.

In the static case where $A$ and $\triangle$ are only functions of $r$ there is a Killing vector that is timelike for $\triangle > 0$. Suitably normalized this Killing vector has components $(1,0,0,0)$ and thus on the Killing horizon, choosing the positive root, equation (\ref{kappasqrd}) gives
\beq \label{surfgrad} \kappa = \frac{A(r_{H})}{4m(r_{H})}\big(1-2m'(r_{H})\big), \eeq
This we will take as the surface gravity of a spherically symmetric Killing horizon. We can now compare how this result for the Killing horizon compares with the various static limits of the proposed dynamical surface gravities in spherically symmetric spacetimes.

The first definition for a dynamical, non-Killing horizon, surface gravity proposed by Hayward in \cite{Hayward:1993wb}, is independent of the chosen normalization on the horizon.
\beq \kappa_{_\mathrm{H}} = \frac{1}{2}\sqrt{-n^{a}\nabla_{a}\theta_{l}}. \eeq
For the spherically symmetric metric we have
\beq n^{a}\nabla_{a}\theta_{l} = \frac{A\triangle -A'r\triangle-Ar^{2}\triangle' }{Ar^{3}}. \eeq
where $m'=\partial_{r}m$. Therefore, on the horizon $\triangle=0$, we have
\beq \kappa_{_\mathrm{H}} = \frac{1}{4m}\sqrt{1-2m'}.\eeq
This will reduce to the static result in situations where $A(r)=1/\sqrt{1-2m'}$. However, as already noted in \cite{Fodor:1996rf} this does not give the correct answer in the Reissner-Nordstr\"{o}m case where $A=1$ and $m'=Q^{2}/2(2M^{2}-Q^{2}+2M\sqrt{M^{2}-Q^{2}})$.\bigskip

Fodor et al.'s value \cite{Fodor:1996rf} for the surface
gravity $\kappa_{\mathrm{F}}$ is simple to evaluate in
Eddington-Finkelstein coordinates. The correctly normalized $n^{a}$ and $l^{a}$ are given by
\beq \label{aEFn} n^{\mu} = \left(0,-1,0,0\right), \eeq
\beq \label{aEFl} l^{\mu} = \left(1,\frac{A\triangle}{2},0,0\right). \eeq
Therefore, we obtain
\beq -n^{a}l^{b}\nabla_{b}l_{a} = \frac{\dot{A}r^{2}+AA'r(r-2m)
+A^{2}(m-m'r)}{Ar^{2}}. \eeq
On the horizon this becomes
\beq \label{Fsurfgrav} \kappa_{_\mathrm{F}} =
\frac{A}{4m}\left(1-2m'\right)+\frac{\dot{A}}{A}. \eeq
This definition will always give the Killing horizon value in
static spacetimes, since in static cases $\dot{A}=0$.\bigskip

The Kodama vector in a spherically symmetric spacetime
has the following form in Eddington-Finkelstein coordinates
\beq K^{\mu} = \left(\frac{1}{A},0,0,0\right). \eeq
%
%
%
%
%
%
%
%
In spherical symmetry we have
\beq g^{ab}K^{c}\left(\nabla_{c}K_{a}-\nabla_{a}K_{c}\right) = \left(\frac{\triangle A' + A\triangle '}{A^{2}},\frac{\triangle (\triangle A' + A\triangle ')}{A},0,0\right) \eeq
on the horizon this becomes
\beq \frac{1}{2}\left(\frac{\triangle '}{A},0,0,0\right) = \kappa_{Ko}K^{\mu} \eeq
and thus we find
\beq \label{Kosurfgrav} \kappa_{_\mathrm{Ko}} = \frac{1}{4m}\left(1-2m'\right), \eeq
which will reduce to the static, Killing horizon case when $A=1$.\bigskip

The value for the surface gravity $\kappa_{\mathrm{M}}$
given by Mukohyama and Hayward in \cite{Mukohyama:1999sp} has the
form
\beq \kappa_{_\mathrm{M}} = -\frac{r}{4}\left(l^{a}\nabla_{a}\theta_{n}+n^{a}\nabla_{a}\theta_{l}+\frac{\theta_{n}}{\Lambda}l^{a}\nabla_{a}\Lambda\right).
\eeq
Since we have
\beq l^{a}\nabla_{a}\theta_{n} =
\frac{2\dot{A}r^{2}+A^{2}\triangle+A'Ar\triangle}{A^{2}r^{3}},
\eeq
this gives
\beq \kappa_{_\mathrm{M}} =
\frac{m}{2}\left(\frac{1-2m'}{4m^{2}}-\frac{\dot{A}}{A^{2}m}\right)-\frac{r\theta_{n}}{4\Lambda}l^{a}\nabla_{a}\Lambda ,
\eeq
Thus the value of the surface gravity depends on knowing the factor $\Lambda$ that makes $l^{a}$ affinely parameterized. This will be discussed further in \ref{doublenull}. In static situations, we have $\Lambda = kA^{2}\triangle$, where $k$ is a constant of integration, and thus
\beq \label{Msurfgrav} \kappa_{_{\mathrm{M}}} = \frac{1}{4M}(1-2m'). \eeq
\bigskip

The surface gravity for a slowly evolving horizon,
proposed by Booth and Fairhurst in \cite{Booth:2003ji}, does not
fix the overall normalization of the surface gravity. The value of
$l^{a}n^{b}\nabla_{b}n_{a}$ depends on the choice of normalization
for $l^{a}$ and for affinely parameterized $n^{a}$ in spherical
symmetry (equation (\ref{aEFn}) above) we naturally find
\beq l^{a}n^{b}\nabla_{b}n_{a} = 0. \eeq
However, for general parameterizations, transforming via $l^{\mu'} = \Omega(x)l^{\mu}$ we find
\beq l^{a'}n^{b}\nabla_{b}n_{a} = \frac{n^{a}\nabla_{a}\Omega}{\Omega^{2}} + \frac{1}{\Omega}l^{a}n^{b}\nabla_{b}n_{a}.
\eeq
Thus, the surface gravity for a slowly evolving horizon depends on
the choice of normalization and will only coincide with that given
by the isolated horizon formula $\kappa =
-n^{a}l^{b}\nabla_{b}l_{a}$ if $n^{a}$ is affinely parameterised or $\Omega' = 0$.

\bigskip

In \cite{Nielsen:2005af} it was shown how the use of the Misner-Sharp mass can lead to a preferred
normalization for the surface gravity in spherically symmetric spacetimes. As shown in \cite{Nielsen:2005af} the surface defined by
\beq \label{surfcond} r=2m(v,r), \eeq
defines a marginally trapped surface at $r=r_{H}$ and in many cases is also a dynamical horizon or trapping horizon.
Differentiating this equation with respect to any parameter $\xi$, labeling spherically symmetric foliations of the horizon, gives
\beq \frac{\d r}{\d\xi} = 2\frac{\partial m}{\partial v}\frac{\d v}{\d\xi} + 2\frac{\partial m}{\partial r}\frac{\d r}{\d\xi}. \eeq
In we take $\xi = v$ and rearrange using the formula for the area $A=4\pi r^2$ this becomes
\beq  \frac{\partial m}{\partial v} = \frac{1}{8\pi}\frac{(1-2m')}{2r}\frac{\d A}{\d v}, \eeq
where $m'=\frac{\partial m}{\partial r}$. In order for this to take the same form as the first law of black hole thermodynamics $\d m = \frac{1}{8\pi}\kappa\;\d A$ it seems natural to take
\beq \label{Nsurfgrav} \kappa_{N} = \frac{(1-2m')}{2r_{H}}. \eeq
This gives a natural normalization of the surface gravity in terms of the Misner-Sharp mass function, which we can interpret as the mass of the black hole contained within the radius $r_{H}$. In the static case, this does not reduce to the usual Killing horizon definition, although this derivation automatically assumes a dynamical evolution since, in the static case, the differentiation with respect to $v$ is not valid.
It is also interesting to note that direct differentiation of (\ref{surfcond}) gives
\beq \frac{\d m}{\d v} = \frac{1}{8\pi}\frac{1}{2r_{H}}\frac{\d A}{\d v}, \eeq
suggesting a surface gravity identical to the one mentioned in \cite{Ashtekar:2004cn} for a dynamical horizon in spherical symmetry.

\section{Conclusion}

We have seen a variety of ways of defining surface gravity for dynamical situations. These seem to fall into three different categories. Firstly there is the very simple form $1/2r_{H}$ in spherical symmetry \cite{Ashtekar:2004cn}. 
The second category is of the form $\kappa=(1-2m')/2m$ (\ref{Kosurfgrav}), (\ref{Msurfgrav}) and (\ref{Nsurfgrav}), while the third category is of the form $A(1-2m')/4m + \dot{A}/A$ (\ref{Fsurfgrav}). Only the third form gives the correct Killing horizon behaviour in the static limit for spherically symmetric spacetimes, although all forms give the correct version for the Schwarzschild spacetime. However, the first and second categories seem to be much more closely related to full dynamical evolution of the horizon. In situations where one can set $A=1$ everywhere, such as the Vaidya solution, the second and third categories will coincide.
  
It thus seems that the issue of the surface gravity is far from clear for general black holes. What role one wants the surface gravity to play depends on the context. In terms of the classical first law of black hole mechanics there is a close connection between the definition of the surface gravity and a choice of quasi-local mass. We have noted that using this approach there is a tension between the surface gravity defined in terms of a Killing horizon and the static limit of the surface gravity defined in terms of the Misner-Sharp quasi-local mass in a spherically symmetric spacetime. Local horizons have many simple properties in spherically symmetric spacetimes. In spherical symmetry the Misner-Sharp mass is equivalent to the Hawking mass and forms a `preferred' quasi-local mass. In rotating spacetimes, such as the Kerr solution, the Hawking mass may not give a useful definition of the mass associated with the horizon \cite{Ashtekar:2004cn} and thus the situation is not as clear.

Furthermore, in spherical symmetry it is easy to show that the Hawking-Gibbons method for calculating the temperature of a static black hole (by eliminating the conical singularity in the Euclidean sector) gives a finite temperature for the Hawking radiation corresponding to the Killing horizon surface gravity (\ref{surfgrad}). However, this version does not correspond to the versions derived from a fully dynamical first law of black hole evolution. Whether this apparent `tension' between the two formulations has any deeper significance for understanding the fully dynamical evolution of black holes remains to be seen.

\section{Acknowledgments}
This work was supported by the ISAT fund of the Royal Society of New Zealand and by the International Cooperation Program of the Korea Science and Engineering Foundation. It is a pleasure to thank Hiromi Saida and Gungwon Kang for helpful discussions and comments.

\appendix

\section{Double null foliations of the Schwarzschild metric}

\label{doublenull}

In \cite{Mukohyama:1999sp} Mukohyama and Hayward define a double
null foliation based on two foliations of null hypersurfaces
labeled by $\xi^{+}$ and $\xi^{-}$. In order to compare their
notation with the notation used here, in the following all tensors
from Mukohyama and Hayward will have a $+$ or a $-$ attached to
them. We will also use abstract index notation for consistency,
although it is not used by Mukohyama and Hayward. They define
one-forms normal to the null hypersurfaces by $n^{\pm} =
-\d\xi^{\pm}$. In abstract index notation this corresponds to
\beq n^{+}_{a} \equiv -\nabla_{a}\xi^{+} \eeq
and
\beq n^{-}_{a} \equiv -\nabla_{a}\xi^{-}. \eeq
Since $\d^{2}\xi^{\pm} = 0$, both of these one-forms are dual to affinely parameterized tangent
vectors of null geodesics. In terms of abstract index notation we have
\beq \nabla_{a}n^{\pm}_{b} - \nabla_{b}n^{\pm}_{a} = 0, \eeq
contracting this with $g^{-1}(n_{\pm})$ we have
\bea n^{\pm a}\left(\nabla_{a}n^{\pm}_{b} -
\nabla_{b}n^{\pm}_{a}\right) & = & 0 \nonumber \\
\Rightarrow n^{\pm a}\nabla_{a}n^{\pm}_{b} -
\frac{1}{2}\nabla_{b}(n^{\pm a}n_{a}^{\pm}) & = & 0 \nonumber \\
\Rightarrow n^{\pm a}\nabla_{a}n^{\pm}_{b} & = & 0, \eea
where the last two lines follow from the Leibniz rule, metric
compatibility of the covariant derivative and the fact that
$n^{\pm a}n_{a}^{\pm} = 0$ everywhere. However, in general we can
choose both null vectors to be non-affinely parameterized by
making the choices
\beq n_{a} = \Gamma n_{a}^{+} \eeq
and
\beq l_{a} = \Lambda n_{a}^{-}, \eeq
where $1/\Gamma$ is a spacetime dependent factor that makes
$n^{a}$ affinely parameterized and $1/\Lambda$ is a spacetime
dependent factor that makes $l^{a}$ affinely parameterized. In
particular, in advanced Eddington-Finkelstein coordinates
$(v,r,\theta,\phi)$, the ingoing null geodesics are naturally
affinely parameterized ($\Gamma =1$) and we can equate
\beq n_{a} = n_{a}^{+} \eeq
and
\beq l_{a} = \Lambda n_{a}^{-}, \eeq
The one-forms have components
\beq n_{\mu} = (-1,0,0,0), \eeq
\beq l_{\mu} = \left(-\frac{A^{2}\triangle}{2},A,0,0\right). \eeq
The corresponding vector components are
\beq n^{\mu} = \left(0,-\frac{1}{A},0,0\right), \eeq
\beq l^{\mu} = \left(1,\frac{A\triangle}{2},0,0\right). \eeq
In this form $n^{a}$ is affinely parameterized while $l^{a}$ is
not and the conventional normalization has been chosen so that
$l^{a}n_{a}=-1$.\bigskip

\noindent Returning to the general case, we can choose the cross-normalization for $l^{a}$ and $n^{a}$ such that
$l^{a}n_{a}=-1$. Since Mukohyama and Hayward have
$g^{-1}(n^{+},n^{-}) = -e^{f}$ we see that
\beq e^{f} = \frac{1}{\Lambda\Gamma}. \eeq
Now Mukohyama and Hayward define two null normal vectors by
$l_{\pm} = e^{-f}g^{-1}(n^{\mp})$ (note the switch as $\pm$
becomes $\mp$.) In our notation this becomes
\beq l_{+}^{a} = \Lambda\Gamma g^{ab}n^{-}_{b} = \Gamma l^{a},
\eeq
\beq l^{a}_{-} = \Lambda\Gamma g^{ab}n^{+}_{b} = \Lambda n^{a}.
\eeq
The expansions of $l_{\pm}$ are defined by $\theta_{\pm} =
*{\cal{L}}_{\pm}*1$ where ${\cal{L}}_{\pm}$ is the Lie derivative
with respect to $l_{\pm}$ and $*$ is the Hodge dual operation
associated with the projection of the metric onto a two-sphere to
which both $l^{+}$ and $l^{-}$ are normal. Using the
anti-symmetric area two-form, or Levi-Civita tensor for the
two-surface, $\epsilon_{ab}$ this can be written
\beq \theta_{\pm} =
\frac{1}{2}\epsilon^{ab}{\cal{L}}_{\pm}\epsilon_{ab} = 0. \eeq
Thus we have
\bea \theta_{-} & = &
\frac{1}{2}\epsilon^{ab}{\cal{L}}_{-}\epsilon_{ab} \nonumber \\
& = &
\frac{\epsilon^{ab}}{2}\left(l^{c}_{-}\nabla_{c}\epsilon_{ab} +
\epsilon_{cb}\nabla_{a}l^{c}_{-} +
\epsilon_{ac}\nabla_{b}l^{c}_{-}\right) \nonumber \\
& = & \frac{\epsilon^{ab}}{2}\left(\Lambda
n^{c}\nabla_{c}\epsilon_{ab} + \epsilon_{cb}\nabla_{a}(\Lambda
n^{c}) +
\epsilon_{ac}\nabla_{b}(\Lambda n^{c})\right) \nonumber \\
& = & \frac{\epsilon^{ab}}{2}\left(\Lambda
n^{c}\nabla_{c}\epsilon_{ab} + \epsilon_{cb}\Lambda\nabla_{a}
n^{c} +
\epsilon_{ac}\Lambda\nabla_{b}n^{c}\right) \nonumber \\
& = & \Lambda\frac{\epsilon^{ab}}{2}{\cal{L}}_{n}\epsilon_{ab},
\eea
where terms such as $\epsilon_{cb}n^{c}\nabla_{a}\Lambda$ are zero
because $n^{c}$ is normal to the surface $\epsilon_{cb}n^{c}=0$.
Similarly we find
$\theta_{+}=\frac{\Gamma}{2}\epsilon^{ab}{\cal{L}}_{l}\epsilon_{ab}$.
The area two-form can be given in terms of the Newman-Penrose null
tetrad, as in Ashtekar et al. \cite{Ashtekar:1999yj}, as
\beq \epsilon_{ab} =
i\left(m_{a}\bar{m}_{b}-\bar{m}_{a}m_{b}\right). \eeq
We can use this to calculate the expansions by
\beq \frac{1}{2}\epsilon^{ab}{\cal{L}}_{n}\epsilon_{ab} =
-\frac{1}{2}(m^{a}\bar{m}^{b}-\bar{m}^{a}m^{b}){\cal{L}}_{n}(m_{a}\bar{m}_{b}-\bar{m}_{a}m_{b}).
\eeq
Performing the necessary permutations, and using the fact that the expansion as
given in (\ref{expanl}) is equivalent to $(m^{a}\bar{m}^{b}+\bar{m}^{a}m^{b})\nabla_{a}l_{b}$ we find
\beq \theta_{-} =
\frac{\Lambda}{2}\epsilon^{ab}{\cal{L}}_{n}\epsilon_{ab} =
\Lambda(m^{a}\bar{m}^{b}+\bar{m}^{a}m^{b})\nabla_{a}n_{b} =
\Lambda \theta_{n} \eeq
and similarly
\beq \theta_{+} =
\frac{\Gamma}{2}\epsilon^{ab}{\cal{L}}_{l}\epsilon_{ab} =
\Gamma(m^{a}\bar{m}^{b}+\bar{m}^{a}m^{b})\nabla_{a}l_{b} =
\Gamma\theta_{l}. \eeq
This gives
\bea
e^{f}\left({\cal{L}}_{+}\theta_{-}+{\cal{L}}_{-}\theta_{+}+\theta_{+}\theta_{-}\right)
& = & l^{a}\nabla_{a}\theta_{n} + n^{a}\nabla_{a}\theta_{l}
\nonumber \\ & & +
\frac{\theta_{n}}{\Lambda}l^{a}\nabla_{a}\Lambda +
\frac{\theta_{l}}{\Gamma}n^{a}\nabla_{a}\Gamma +
\theta_{n}\theta_{l}.\eea
The surface gravity of Mukohyama and Hayward, for a spherically
symmetric trapping horizon is
\bea \kappa_{_\mathrm{M}} & = & -\frac{1}{16\pi
r}\int_{S}\d^{2}\theta\sqrt{h}e^{f}\left({\cal{L}}_{+}\theta_{-}+{\cal{L}}_{-}\theta_{+}+\theta_{+}\theta_{-}\right)
\nonumber \\ & = &
-\frac{r}{4}\left(l^{a}\nabla_{a}\theta_{n}+n^{a}\nabla_{a}\theta_{l}+\frac{\theta_{n}}{\Lambda}l^{a}\nabla_{a}\Lambda\right).
\eea
In order to use this definition we need to know what the factor
$\Lambda$ is that makes $l^{a}$ affinely parameterized. To do this we
need to pick a coordinate system that is regular on the horizon.
It is difficult, at least in advanced Eddington-Finkelstein
coordinates, to calculate what the affine parameter for the
outgoing null vectors should be. By examining the way the surface
gravity changes under a reparameterization of $l^{a}$ we see that
for a non-affine parameterization we have
\beq l^{a}\nabla_{a}\Lambda = -\Lambda
n^{a}l^{b}\nabla_{b}l_{a}.\eeq
Using the non-affine parameterization given above, (\ref{aEFn}) and (\ref{aEFl}), this becomes
\beq \partial_{v}(\Lambda) +
\frac{A\triangle}{2}\partial_{r}(\Lambda) =
\Lambda\left(\frac{\dot{A}}{A}+A'\triangle +
\frac{A\triangle'}{2}\right) \eeq
or
\beq \partial_{v}(\mathrm{ln}\Lambda) +
\frac{A\triangle}{2}\partial_{r}(\mathrm{ln}\Lambda) =
\frac{\dot{A}}{A}+A'\triangle + \frac{A\triangle'}{2} \eeq
The solution of this first order partial differential equation will depend on the specific situation. For static solutions we have
$\dot{A}=0=\partial_{v}(\mathrm{ln}\Lambda)$ and the solution to
the ordinary differential equation is just
\beq \Lambda = kA^{2}\triangle, \eeq
where $k$ is a constant of integration. Therefore, for the static
case we have $l^{a}\nabla_{a}\theta_{n}=0$,
$n^{a}\nabla_{a}\theta_{l} = -\triangle'/r$, $\theta_{n} =
-2/(Ar)$ and
\beq \kappa_{_{\mathrm{M}}} = \frac{1}{4M}(1-2m'), \eeq
which we see will match the Killing horizon value for $A=1$. In
particular, in the static Schwarzschild solution $\Lambda =
\triangle$ and we have, at the horizon $\triangle=0$,
\beq n^{a}\nabla_{a}\theta_{l} = -\frac{2M}{r^{3}} \eeq
\beq l^{a}\nabla_{a}\theta_{n} = 0 \eeq
\beq \frac{\theta_{n}}{\Lambda}l^{a}\nabla_{a}\Lambda =
-\frac{2M}{r^{3}} \eeq
\beq \kappa_{_\mathrm{M}} = \frac{M}{r^{2}} = \frac{1}{4M}. \eeq
\bigskip

\noindent In Kruskal-Szekeres coordinates ($T,R,\theta,\phi$) we
can easily generate a double null foliation and thus find affinely
parameterized null normals. The Schwarzschild solution takes the
form
\beq \d s^{2} = -\frac{32M^{3}}{r}e^{-r/2M}(-\d T^{2} + \d R^{2})
+ r^{2}\d\Omega^{2}. \eeq
This can be written in double null coordinates by identifying
\beq v = T + R, \eeq
\beq u = T - R. \eeq
In double null coordinates $(u,v,\theta,\phi)$ the Schwarzschild
metric takes the form
\beq \d s^{2} = -\frac{32M^{3}}{r}e^{-r/2M}\d u\d v +
r^{2}\d\Omega^{2}. \eeq
The radial null vectors are
\beq l^{\mu} = \left(0,\frac{r}{16M^{3}}e^{r/2M},0,0\right), \eeq
\beq l_{\mu} = (-1,0,0,0), \eeq
and
\beq n^{\mu} = \left(\frac{r}{16M^{3}}e^{r/2M},0,0,0\right), \eeq
\beq n_{\mu} = (0,-1,0,0). \eeq
For these choices we have $l^{a}l_{a}=0=n^{a}n_{a}$ and
\beq l^{a}n_{a} = -\frac{r}{16M^{3}}e^{r/2M}. \eeq
Thus, if we take $l^{a}$ to be $l_{+}$ and $n^{a}$ to be $n_{-}$,
in terms of Mukohyama and Hayward's notation, we have
\beq e^{f} = \frac{r}{16M^{3}}e^{r/2M}, \eeq
\beq \theta_{+} = \frac{e^{r/2M}\partial_{v}r}{8M^{3}}, \eeq
\beq \theta_{-} = \frac{e^{r/2M}\partial_{u}r}{8M^{3}}, \eeq
\beq e^{f}{\cal{L}}_{-}\theta_{+} =
-\frac{e^{r/2M}\left[\partial_{v}r\partial_{u}r -
r\partial_{u}\partial_{v}r\right]}{8M^{3}r}, \eeq
\beq e^{f}{\cal{L}}_{+}\theta_{-} =
-\frac{e^{r/2M}\left[\partial_{v}r\partial_{u}r -
r\partial_{u}\partial_{v}r\right]}{8M^{3}r}. \eeq
The surface gravity for the horizon is now simply given by
\beq \kappa_{_{\mathrm{M}}} =
-\frac{r}{4}e^{f}\left({\cal{L}}_{-}\theta_{+}+{\cal{L}}_{+}\theta_{-}\right).
\eeq
Since we can write the double null coordinates $u,v$ in terms of
Schwarzschild coordinates $r,t$ as
\beq v = \sqrt{\frac{r}{2M}-1}e^{(r+t)/4M}, \eeq
\beq u = -\sqrt{\frac{r}{2M}-1}e^{(r-t)/4M}, \eeq
we have
\beq uv = \left(1-\frac{r}{2M}\right)e^{r/2M} \eeq
and thus, at the horizon $r=2M$
\beq \partial_{u}\partial_{v}r = - 2M. \eeq
Putting it all together with $\theta_{+}=0 \Rightarrow
\partial_{v}r=0$ gives
\beq \kappa_{_{\mathrm{M}}} = \frac{1}{4M}. \eeq
This value agrees with the value given by the Killing horizon
method for the Schwarzschild solution.

\end{document}